\documentclass[twocolumn,prd, superscriptaddress,preprintnumbers,nofootinbib]{revtex4}
\usepackage{graphicx}
\usepackage{amsmath}
\usepackage{amssymb}
\usepackage{color}
\usepackage{ulem}
\usepackage{hyperref}

\newcommand{\beq}{\begin{equation}}
\newcommand{\eeq}{\end{equation}}
\newcommand{\bea}{\begin{eqnarray}}
\newcommand{\eea}{\end{eqnarray}}

\def\pT{p_{\text{T}}}

\def\met{\slash{\!\!\!\! E}_{\text{T}}}
\allowdisplaybreaks[1]

\begin{document}

\title{A New Observable for Measuring CP Property of Top-Higgs Interaction}

\author{Qing-Hong Cao}
\email{qinghongcao@pku.edu.cn}
\affiliation{Department of Physics and State Key Laboratory of Nuclear Physics and 
Technology, Peking University, Beijing 100871, China}
\affiliation{Collaborative Innovation Center of Quantum Matter, Beijing 100871, China}
\affiliation{Center for High Energy Physics, Peking University, Beijing 100871, China}

\author{Ke-Pan Xie}
\email{kpxie@snu.ac.kr}
\affiliation{Center for Theoretical Physics, Department of Physics and Astronomy, 
Seoul National University, Seoul 08826, Korea}

\author{Hao Zhang}
\email{zhanghao@ihep.ac.cn}
\affiliation{Theoretical Physics Division, Institute of High Energy Physics, Beijing 100049, China}
\affiliation{School of Physics, University of Chinese Academy of Science, Beijing 100049, China}
\affiliation{Center for High Energy Physics, Peking University, Beijing 100871, China}

\author{Rui Zhang}
\email{rui.z@pku.edu.cn}
\affiliation{Department of Physics and State Key Laboratory of Nuclear Physics and 
Technology, Peking University, Beijing 100871, China}

\begin{abstract}
We propose a new dihedral angle observable to measure the CP property of the interaction of top quark and Higgs boson in the $t\bar{t}H$ production at the 14~TeV LHC. We consider two decay modes of the Higgs boson, $H\to b\bar{b}$ and $H\to \gamma\gamma$ and show that  the dihedral angle distribution is able to distinguish the CP-even and the CP-odd hypothesis at 95\% confidence level with an integrated luminosity of $\sim 180~{\rm fb}^{-1}$.
\end{abstract}

\maketitle

\noindent{\bf 1. Introduction}
\vspace*{2mm}

In the standard model (SM) of particle physics, the Higgs boson is a CP-even scalar boson with $J^{PC}=0^{++}$. Any deviation from this prediction is a clear evidence of new physics (NP) beyond the SM. Therefore, measuring the CP nature of the Higgs boson is a hot topic at the Large Hadron Collider (LHC)~\cite{Aad:2015mxa,Aad:2020mnm,Sirunyan:2019twz,Sirunyan:2019nbs}. The interaction between top quark and Higgs boson has been verified in the $t\bar{t}H$ channel recently~\cite{Sirunyan:2018hoz,Aaboud:2018urx}, and the next target is to measure the CP property of the $Ht\bar{t}$ interaction in the $t\bar{t}H$ channel~\cite{Sirunyan:2020sum,Aad:2020ivc}. The effective Lagrangian of the $Ht\bar{t}$ interaction can be parameterized as 
\beq
\mathcal L=-Y_t\bar t e^{i\alpha\gamma_5}tH~~~~\alpha\in[0,2\pi),~~Y_t\in \mathbb R^+,
\label{eq:3}
\eeq
with $\alpha$ denotes the CP-phase angle. Many observables and methods have been proposed in the literature~\cite{Gunion:1996xu,Boudjema:2015nda,Mileo:2016mxg,Gritsan:2016hjl,AmorDosSantos:2017ayi,Santos:2015dja,Gouveia:2018xfp,Goncalves:2018agy,Ren:2019xhp,Gouveia:2019trd,Ferroglia:2019qjy,Bahl:2020wee,Bortolato:2020zcg}, and most of them require fully reconstructing the kinematics of both the top quark and antitop quark, which is very challenging. In this work, we propose a novel observable which demands reconstructing only one top quark. 

\begin{figure}
\includegraphics[scale=0.3]{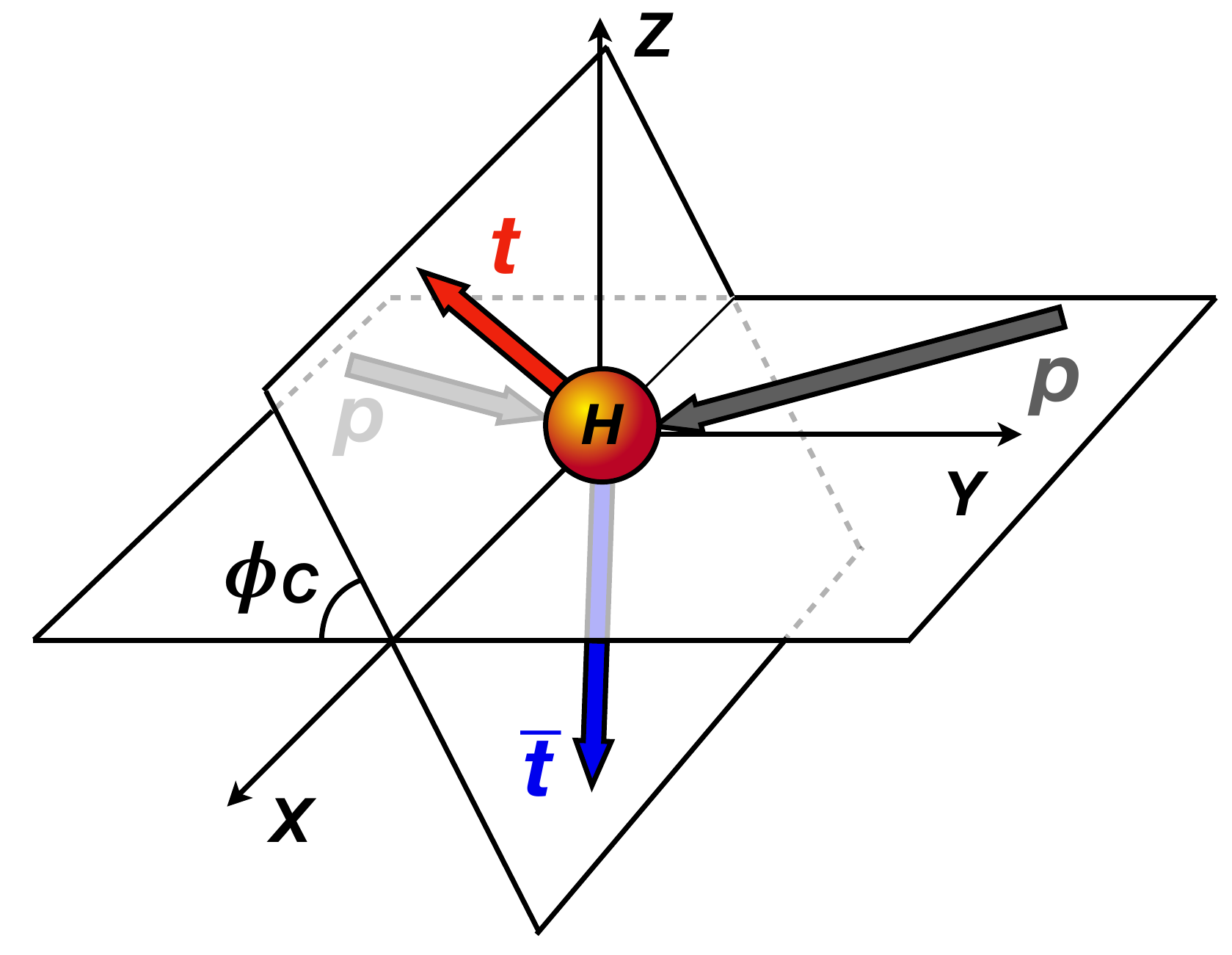}
\caption{The dihedral angle $\phi_C$ between the plane of incoming protons and the plane of the $t\bar{t}$ pair in the rest frame of the Higgs boson. }
\label{fig1}
\end{figure}

The observable is a dihedral angle ($\phi_C$) between the plane spanned by the incoming protons and the plane spanned by the $t\bar t$ pair in the rest frame of Higgs boson, as depicted in Fig.~\ref{fig1}. 
The head-on collision $pp\to t\bar{t}H$ in the laboratory frame can be viewed approximately as a non-head-on ``$2\to 2$" scattering in the rest frame of the Higgs boson, e.g. the two colliding protons produce two moving top quarks and one Higgs boson at rest. In such a picture the non-zero 3-momentum of the incoming parton pair is equal to that of the top quark pair in the final state while the Higgs boson merely carries away a rest energy. 

\begin{figure*}
\begin{center}
\includegraphics[scale=0.4]{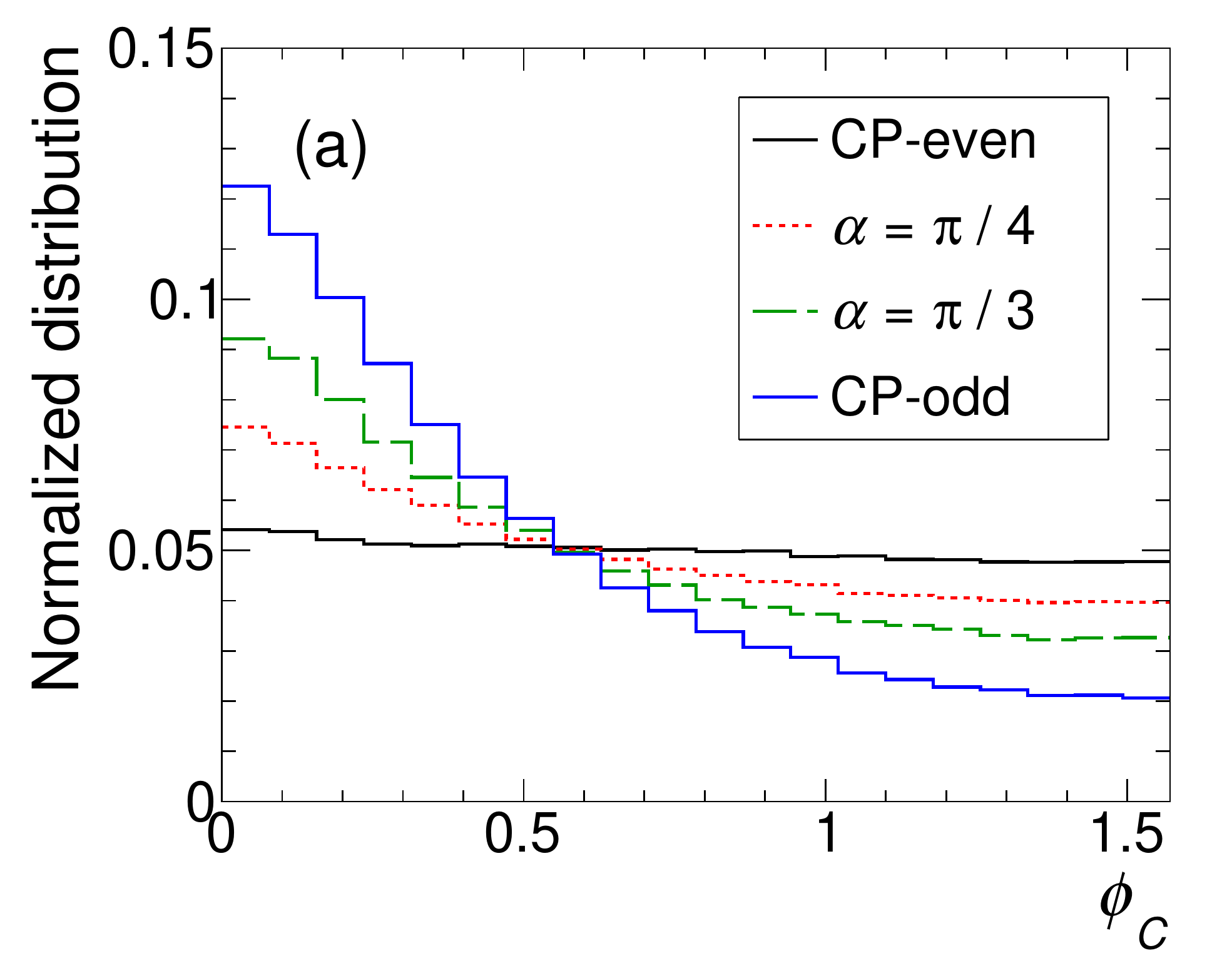}
\includegraphics[scale=0.4]{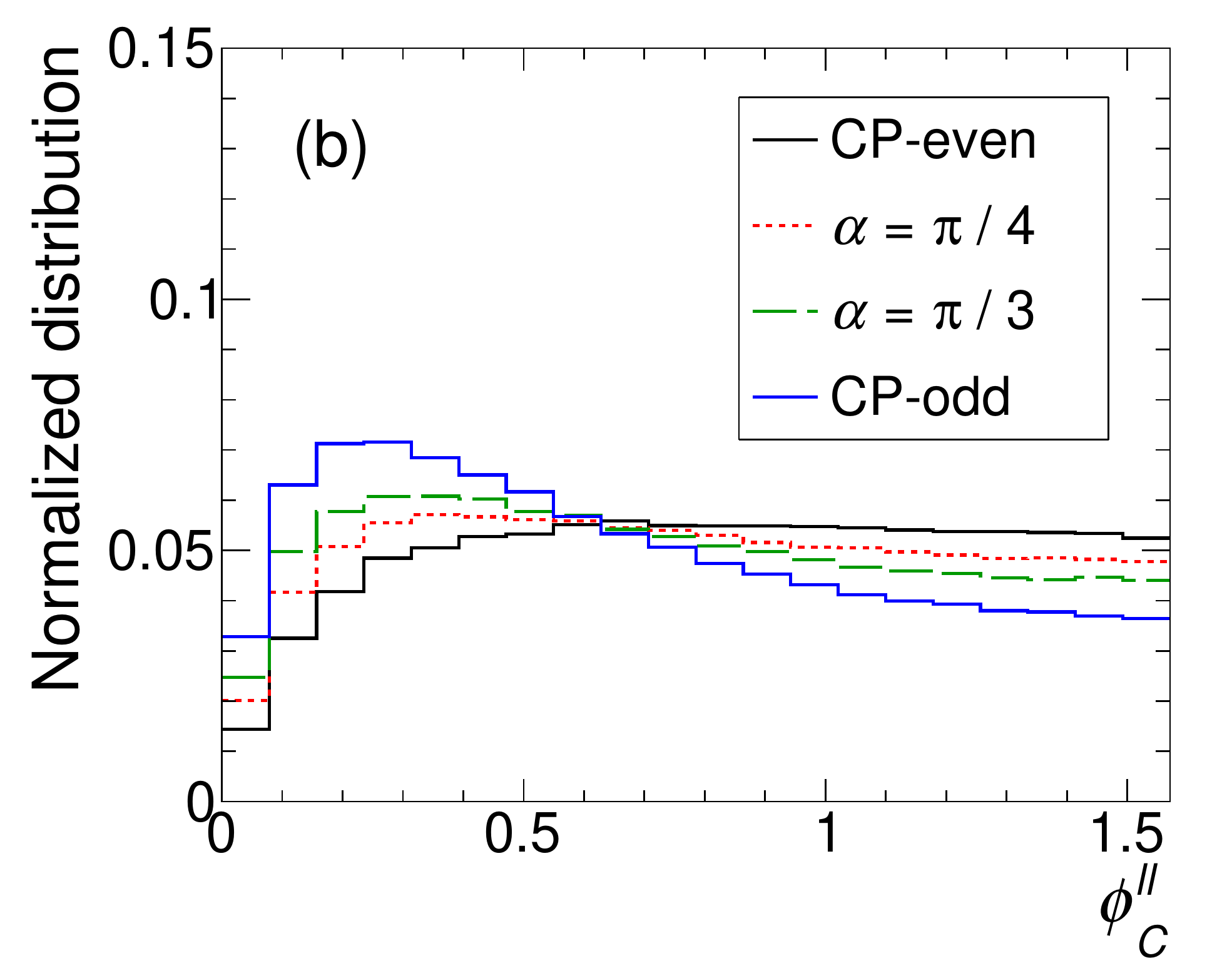}
\caption{Normalized distributions of $\phi_C$ (a) and $\phi_C^{\ell\ell}$ (b) for various CP phase angles: $\alpha=0$ (CP-even), $\pi/4$, $\pi/3$, and $\pi/2$ (CP-odd)}
\label{fig2}
\end{center}
\end{figure*}

Denote the normalized 3-momenta of the protons, top quark and antitop quark in the Higgs rest frame as ${\bf n}_{p_1},{\bf n}_{p_2},{\bf n}_{t}$ and ${\bf n}_{\bar t}$, respectively, the cosine of the dihedral angle is 
\beq
\cos\phi_C=\frac{\left|({\bf n}_{p_1}\times{\bf n}_{p_2})\cdot({\bf n}_{t}\times{\bf n}_{\bar t})\right|}{\left|{\bf n}_{p_1}\times{\bf n}_{p_2}\right|\cdot\left|{\bf n}_{t}\times{\bf n}_{\bar t}\right|}.
\label{eq:phic}
\eeq
Without loss of generality, we choose the $XY$-plane as the plane of the incoming protons and the positive $X$-axis along the direction of the total 3-momenta of the incoming protons. The $Z$-axis is chosen such that $\vec{p}_z^{~t}>0$. As the two protons are identical, it is meaningless to distinguish $\phi_C$ and $\pi-\phi_C$; therefore, we restrict the range of the $\cos\phi_C$ in $[0,1]$. 
Figure~\ref{fig2}(a) displays the normalized $\phi_C$ distributions at the14 TeV LHC for four benchmark CP phase angles $\alpha$'s, e.g. $\alpha=0$ (CP-even), $\pi/4$, $\pi/3$, and $\pi/2$ (CP-odd). Note that the possibility of the Higgs boson being a purely CP-odd scalar is fading away after considering various Higgs boson production channels~\cite{Cao:2015oaa,Chen:2015rha,Cao:2016zob,Cao:2016wib,Cao:2019ygh}. The simulation is done by using MadGraph5 \cite{Frederix:2018nkq} with CT14llo parton distribution function (PDF) \cite{Dulat:2015mca}. While the CP-odd Higgs-Top interaction exhibits a peak in the small $\phi_C$ region,  the CP-even coupling has a flat distribution. The difference can be used to measure the phase angle $\alpha$. 

To suppress the SM background, the dileptonic decay mode of $t\bar t$ in the final state is often used. Unfortunately, the reconstruction of the (anti)top quark kinematics is challenging in the case. Because the charged lepton from the (anti)top quark decay is maximally correlated with the spin of (anti)top quark~\cite{Czarnecki:1990pe, Brandenburg:2002xr,Cao:2004ky,Cao:2007ea,Heim:2009ku}, we define the dihedral angle between the plane of two charged leptons and the plane of incoming protons as following:
\beq
\cos\phi_C^{\ell\ell}=\frac{\left|({\bf n}_{p_1}\times{\bf n}_{p_2})\cdot({\bf n}_{\ell^+}\times{\bf n}_{\ell^-})\right|}
{\left|{\bf n}_{p_1}\times{\bf n}_{p_2}\right|\cdot\left|{\bf n}_{\ell^+}\times{\bf n}_{\ell^-}\right|}.
\label{eq:phi_ll}
\eeq
Figure~\ref{fig2}(b) displays the $\phi_C^{\ell\ell}$ distributions for the four CP phases.  The $\phi_C^{\ell\ell}$ distribution is distorted in the small angle region but still can be used to discriminate the CP properties of the $Ht\bar{t}$ interaction.

\noindent{\bf 2. Collider Simulation}
\vspace*{2mm}

The $\alpha$-dependence of the $t\bar{t}H$ production cross section at the leading order (LO) at 14 TeV LHC can be parameterized as  
\beq
\sigma(\alpha)_{pp\to t\bar tH}=0.216\sin^2\alpha+0.484\cos^2\alpha~({\rm pb}).
\label{eq:4}
\eeq
We perform a fast collider simulation at the parton level to demonstrate the potential of the dihedral angles, $\phi_C$ and $\phi_C^{\ell\ell}$, in the measurement of the CP phase of the $Ht\bar{t}$ interaction. Since the dihedral angles are defined in the rest frame of the Higgs boson, it is important to reconstruct the full kinematics of the Higgs boson. For that we focus on the $H\to b\bar b$ and $H\to \gamma\gamma$ decay modes of the Higgs boson. Furthermore, we only consider the dominant SM backgrounds. Our cut-based parton-level analysis demonstrates that the dihedral angle distributions are good at measuring the CP phase $\alpha$ such that it can be used to expedite the BDT method. 

We generate the signal and background events at the LO using MadGraph5~\cite{Frederix:2018nkq} with the CT14llo PDF~\cite{Dulat:2015mca}. The $t\bar tH$ production rate is rescaled such that the total cross section for the CP-even Higgs case is the NLO cross section which includes both the QCD and EW corrections~\cite{deFlorian:2016spz}. To mimic the detector effects, we introduce  Gaussian smearing effects in the transverse momentum ($p_T$) of charged leptons, jets and photons as follows:
\bea
\frac{\sigma_{e^\pm,\gamma}}{\pT}&=&\left\{\begin{array}{cc}
0.0013\oplus\displaystyle{\frac{0.03}{\sqrt{\pT/{\text{GeV}}}}} & |\eta|\leqslant 0.5, \\
0.0017\oplus\displaystyle{\frac{0.05}{\sqrt{\pT/{\text{GeV}}}}} & 0.5<|\eta|\leqslant 1.5, \\
0.0031\oplus\displaystyle{\frac{0.15}{\sqrt{\pT/{\text{GeV}}}}} & 1.5<|\eta|\leqslant 2.47,
\end{array}\right.\nonumber\\
\frac{\sigma_{\mu^\pm}}{\pT}&=&\left\{\begin{array}{cc}
0.0001\oplus\displaystyle{\frac{0.01}{\sqrt{\pT/{\text{GeV}}}}} & |\eta|\leqslant 0.5, \\
0.00015\oplus\displaystyle{\frac{0.015}{\sqrt{\pT/{\text{GeV}}}}} & 0.5<|\eta|\leqslant 1.5, \\
0.00035\oplus\displaystyle{\frac{0.025}{\sqrt{\pT/{\text{GeV}}}}} & 1.5<|\eta|\leqslant 2.5,
\end{array}\right.\nonumber\\
\frac{\sigma_{j,b}}{\pT}&=&0.06\oplus\frac{0.95}{\sqrt{\pT/{\text{GeV}}}}.
\eea
The $b$-tagging efficiency is chosen as 80\% while the rate of a charm-jet faking a $b$-jet is chosen as 10\% and the fake-rate of a light-jet is 1\%. 

\vspace*{6mm}
\noindent{\it 2.1 The $H\to \gamma\gamma$ mode}
\vspace*{2mm}

In this channel, in order to keep more signal events, we require the semileptonic decay mode of the $t\bar t$ in the final state, i.e. $t\bar{t}\to b\bar{b}jj\ell^\pm\nu$. The event topology of the signal events consists of one isolated charged lepton ($e^\pm$ or $\mu^\pm$), two $b$-tagged jets, two photons arising from the Higgs boson decay, two non-$b$-tagged jets and large missing transverse energy from the invisible neutrino. The dominant SM background is from the channel of $pp\to t\bar t\gamma\gamma$ while other backgrounds, e.g., $pp\to VV jj \gamma\gamma$, are sub-dominant.  

We impose a set of pre-selection cuts as follows:
\bea
&&\pT^b>40~{\text{GeV}},~~|\eta^b|<2.5,~~\pT^j>25~{\text{GeV}},~~|\eta^j|<4.5,\nonumber\\
&&\pT^{\ell^\pm}>15~{\text{GeV}},~~|\eta^{\ell^\pm}|<2.4,~~\met>40~{\text{GeV}},\nonumber\\
&&E_{\text{T}}^{{\text{leading}}~\gamma}>35~{\text{GeV}}, ~~E_{\text{T}}^{{\text{subleading}}~\gamma}>25~{\text{GeV}},\nonumber\\
&&|\eta^{\gamma}|<2.4,~~\Delta R_{ik}>0.4,~i,k=b,\ell^\pm,j,\gamma,\nonumber\\
&&|m_{\gamma\gamma}-m_H|<5~{\text{GeV}},
\label{eq:cut0}
\eea
where $\Delta R_{ik}$ is the angular distance between the objects $i$ and $k$, defined as 
\beq
\Delta R_{ik}=\sqrt{(\eta_i-\eta_k)^2 + (\phi_i -\phi_k)^2},
\eeq
and $m_H$ denotes the mass of the Higgs boson, which is chosen as $m_H=125~{\rm GeV}$ throughout this work. Assuming the $j\to \gamma$ fake-rate being $10^{-5}$, we find that the cross sections of the background processes of $t\bar t\gamma j$, $t\bar t jj$ and $VVjj\gamma\gamma$ are about $10^{-4}~{\rm fb}$ after the pre-selection cuts and are ignored in our analysis. 

It is straightforward to reconstruct the kinematics of the Higgs boson from the two energetic photons. Furthermore, we demand three cuts, based on the property of top quark decays, as follows:
\bea
&&|m_{jj}-80~{\text{GeV}}|<20~{\text{GeV}},\nonumber\\
&&|m_{bjj}-175~{\text{GeV}}|<25~{\text{GeV}},\nonumber\\
&&m_{b\ell}<140~{\text{GeV}}.
\eea
to suppress the backgrounds. The likelihood fitting method is used to pick up the correct combinations of those jets from the $W$-boson decay and the top quark decay. We fit the invariant mass distributions of the $(b\ell)$, $(\ell\nu)$, $(b\ell\nu)$,  $(jj)$ and $(bjj)$ systems using the likelihood functions as follows: 
\bea
L_{b\ell}(m)&=&\displaystyle{\frac{m}{(130.1)^2~{\rm GeV}}}\left[1+\left(\frac{m}{63.8}\right)^2\right]\biggl\{1-\nonumber\\
&&\tanh^2\biggl[\displaystyle{\frac{m}{149.0}}+\left(\frac{m}{149.0}\right)^6+\left(\displaystyle{\frac{m}{179.0}}\right)^{12}\biggr]\biggr\},\nonumber\\
L_{\ell\nu}(m)&=&\displaystyle{\frac{1}{(7.5~{\text{GeV}})\pi\left[1+\left(\displaystyle{\frac{m-81.4}{7.5}}\right)^2\right]}},\nonumber\\
L_{b\ell\nu}(m)&=&\displaystyle{\frac{1}{(13.1~{\text{GeV}})\pi\left[1+\left(\displaystyle{\frac{m-174.7}{13.1}}\right)^2\right]}},\nonumber\\
L_{jj}(m)&=&\displaystyle{\frac{1}{\sqrt{2\pi}\times8.3~{\text{GeV}}}}\exp\left[-\displaystyle{\frac{1}{2}\frac{(m-81.0)^2}{(8.3)^2}}\right],\nonumber\\
L_{bjj}(m)&=&\displaystyle{\frac{1}{\sqrt{2\pi}\times13.6~{\text{GeV}}}}\exp\left[-\displaystyle{\frac{1}{2}\frac{(m-174.7)^2}{(13.6)^2}}\right],~~~~
\label{eq:bl}
\eea
where the parameter $m$ is in the unit of GeV. Minimizing the following logarithm of likelihood function (LL)
\beq
-2\log L_{b\ell}-2\log L_{b\ell\nu}-2\log L_{\ell\nu}-2\log L_{jj}-2\log L_{bjj}\nonumber
\eeq
with the $Z$-direction component of the neutrino $p_z^\nu$ as a variable, we determine which $b$-jet is from the leptonic decaying (anti-)top quark and also solve the $p_z^\nu$ simultaneously. The cross sections of the signal and dominant SM background after pre-selection cuts and reconstruction are shown in  Table~\ref{tab1}. The number of the signal events after event reconstruction is small due to the small branching ratio ${\text{Br}}(H\to \gamma\gamma)$.

\begin{table}
\caption{The cross section (in the unit of fb) of the signal process ($\alpha=0$ and 
$\alpha=\pi/2$) and the major background process $t\bar{t}\gamma\gamma$ in the semileptonic mode of the top quark pair.}
\begin{center}
\begin{tabular}{l|c|c|c}
\hline\hline
&~~~$\alpha=0$~~~&~~~$\alpha=\pi/2$~~~&~~~$t\bar t\gamma\gamma$~~~\\
\hline\hline
~~After pre-selection cuts~~&~~$0.0345$~~&~~$0.0140$~~&~$0.0056$~\\
~~After reconstruction~~&~~$0.0189$~~&~~$0.0074$~~&~$0.0029$~\\
\hline\hline
\end{tabular}
\end{center}
\label{tab1}
\end{table}

\begin{figure}[htb]
\begin{center}
\includegraphics[width=0.45\textwidth]{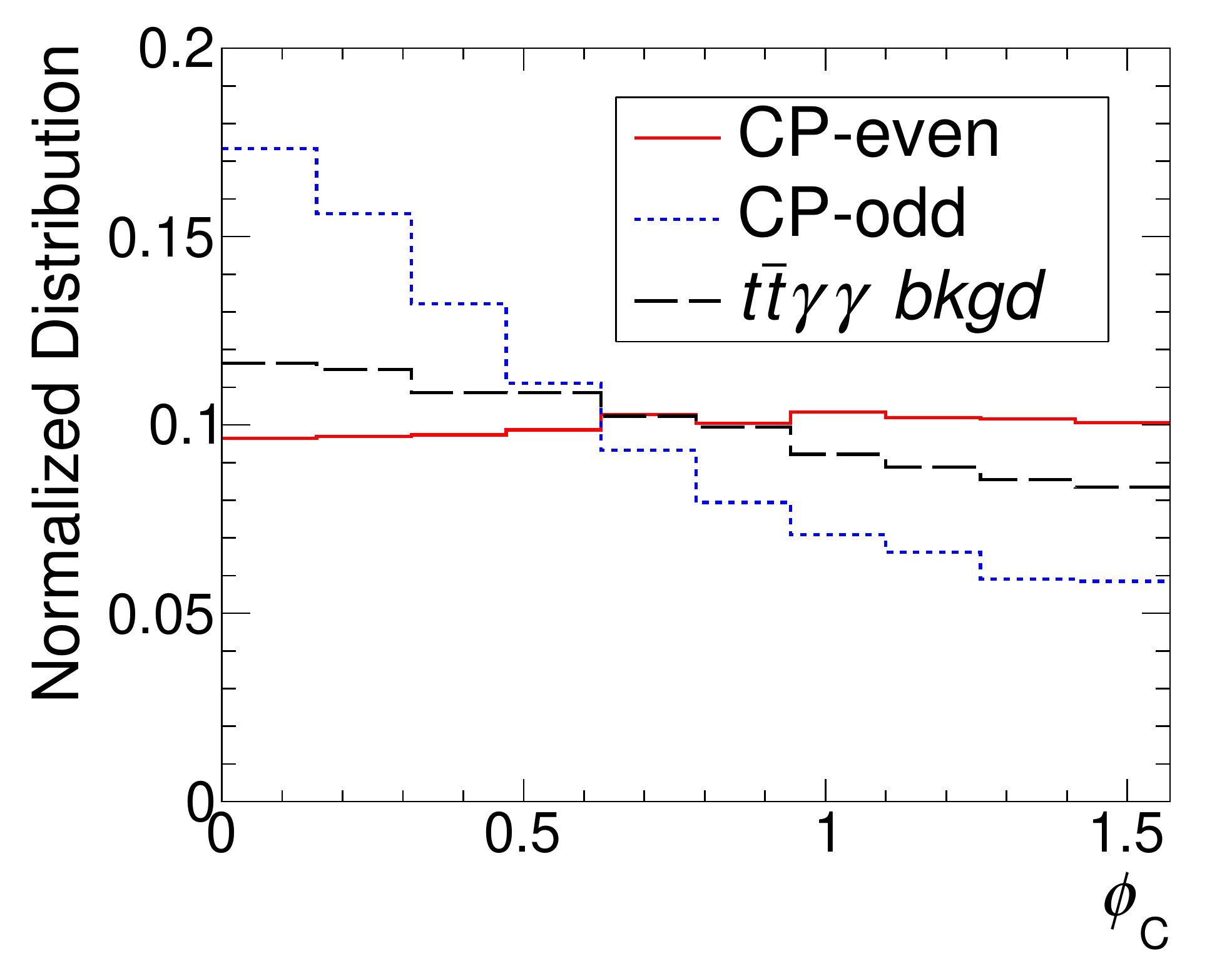}
\caption{The normalized $\phi_C$ distribution in the $pp\to t\bar tH\to \gamma\gamma\ell^\pm \bar bbjj+
\displaystyle{\not{\!\!E}_{\text{T}}}$ channel after the event reconstruction.}
\label{fig4}
\end{center}
\end{figure}

Once the full kinematics of the top quark and the Higgs boson are reconstructed, we calculate the $\phi_C$ angle, defined in Eq.~(\ref{eq:phic}). The normalized $\phi_C$ distributions is plotted in Fig.~\ref{fig4}. The difference between the CP-even and CP-odd Higgs bosons still remains after the event reconstruction. 

\vspace*{6mm}
\noindent{\it 2.2 The $H\to b\bar b$ mode}
\vspace*{2mm}

To suppress the SM background, we consider the dileptonic decaying mode of $t\bar t$, i.e., $t\bar{t}\to b\bar{b}\ell^+\ell^-\nu\bar{\nu}$. The dominant SM background is $pp\to t\bar tb\bar b$. The event topology of the signal contains two opposite-sign charged leptons ($e^\pm$ or $\mu^\pm$), four $b$-tagged jets, and large missing transverse momentum. In order to select the signal event, we impose a set of pre-selection cuts as follows:
\bea
&&\pT^b>40~{\text{GeV}},~|\eta^b|<2.5,~\pT^{\ell^\pm}>20{\text{GeV}},~|\eta^{\ell^\pm}|<2.4,\nonumber\\
&&\Delta R_{ik}>0.4~(i,k=b,\ell^\pm),~~~\met>50~{\text{GeV}}.
\eea
When the two charged leptons are of the same flavor, e.g. $e^+e^-$ or $\mu^+\mu^-$, we require they are away from the $Z$ pole, i.e.   
\beq
|m_{\ell^+\ell^-}-m_Z|>10~{\text{GeV}},
\eeq
to suppress the $Z+{\text{jets}}$ background. In addition, we require $m_{\mu^+\mu^-}>20$ GeV to suppress the background from heavy flavor hadron decay.

When the two $b$-jets are from the Higgs boson decay, their invariant mass must peak around $m_h$; therefore, we require at least one pair of $b$-jets satisfying the following invariant mass cut, 
\beq
|m_{bb}-m_H|<25~{\text{GeV}}.
\eeq
The other two $b$-jets and two charged leptons are from the top quark decay. The invariant mass of the $b$-jet and the charged lepton, if they originate from the same top quark decay, is less than 140~GeV, owing to the spin correlation effect.

For event reconstruction it is crucial to determine which two $b$-jets from the Higgs boson decay, which is done with the likelihood fitting method in our analysis. The likelihood function of the invariant mass of the $b\bar b$ pair from the Higgs boson decay is 
\bea
L_{bb}(m)&=&\displaystyle{\frac{1}{\sqrt{2\pi}\times10.6~{\text{GeV}}}}\exp\left[-\displaystyle{\frac{1}{2}\frac{(m-126.2)^2}{(10.6)^2}}\right],~~~~~~
\eea
after imposing all the cuts. Again, the parameter $m$ is in the unit of GeV. The $b\ell^\pm$ distributions are used to decrease the contamination from the $b$-jets from the top quark decay. We demand any pair of the $b$-jet and the charged leptons must satisfy the following condition, 
\beq
m_{b\ell}<140~{\text{GeV}},
\eeq
and then fit the invariant mass distributions of the $b\ell^\pm$ pair with the likelihood function $L_{b\ell}$ given in Eq.~(\ref{eq:bl}). By minimizing the discriminator, 
\beq
D=-22.0-5\log L_{bb}-0.02\sqrt{\log^2L_{b\ell^+}+\log^2L_{b\ell^-}},~~\nonumber
\eeq
we identify the two $b$-jets from the Higgs boson decay. In addition, a cut of $D<0$ is imposed to optimize the signal-to-background ratio. 

Table~\ref{tab2} shows the cross section of the signal ($\alpha=0$ and $\alpha=\pi/2$) and dominant SM backgrounds after the pre-selection cut and the event reconstruction.  The rate of other backgrounds, e.g. $W^+W^-+4j$, $W^+W^-+1b3j$, $W^+W^-+2b2j$ and $W^+W^-+3b1j$, are smaller than $10^{-5}~\rm{fb}$ after the pre-selection cuts and are ignored in our analysis.  

\begin{table}
\caption{The cross section (in the unit of fb) of signal and background processes where $j$ denotes the light-flavor jet from $g,u,d,s,c$. }
\begin{center}
\begin{tabular}{c|c|c|c|c|c|c}
\hline\hline
&~$\alpha=0$~&$\alpha=\pi/2$&$t\bar tb\bar b$&
$t\bar tbj$&$t\bar tjj$&$WW4b$\\
\hline\hline
pre-selection& 0.601 & 0.295 & 1.261 & 0.0215 & 0.0460 & 0.0007\\
reconstruction& 0.558 & 0.273 & 0.945 & 0.0160 & 0.0343 & 0.0005 \\
\hline\hline
\end{tabular}
\end{center}
\label{tab2}
\end{table}

\begin{figure}
\begin{center}
\includegraphics[width=0.45\textwidth]{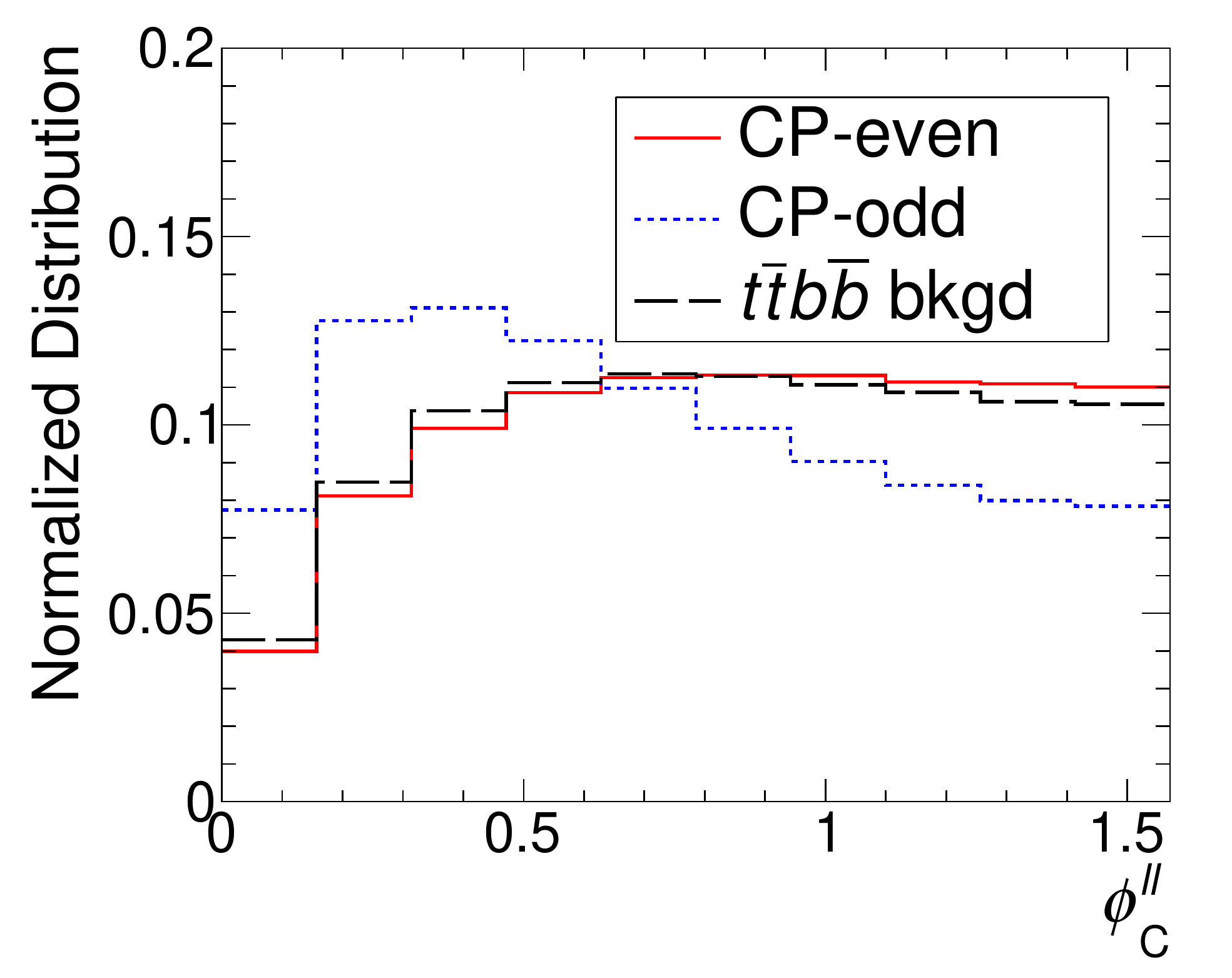}
\caption{The normalized $\phi_C^{\ell\ell}$ distribution the $pp\to t\bar tH
\to 4b+\ell^+\ell^-+\met$ channel. }
\label{fig5}
\end{center}
\end{figure}

After identifying the two $b$-jets from the Higgs boson decay, the other two $b$-jets are treated as from the top quark decays. Owing to the two invisible neutrinos in the final state, it is hard to reconstructed the top quark and antitop quark. We consider the $\phi_C^{\ell\ell}$ defined in Eq.~(\ref{eq:phi_ll}) and plot the normalized distributions in Fig.~\ref{fig5}. The CP-even Higgs boson (red) and the SM background (black) share almost the same distribution. On the other hand, the CP-odd Higgs boson (black curve) exhibits a distinct distribution. 

\vspace*{6mm}
\noindent{\bf 3. CP-even versus CP-odd}
\vspace*{2mm}

A purely CP-odd scalar is severely limited by the global fitting of the single Higgs boson production, the $t\bar{t}H$ production and the $t\bar{t}t\bar{t}$ production~\cite{Chen:2015rha,Cao:2019ygh,Cao:2016wib}. It is still important to probe the CP phase directly from a single scattering process. Equipped with the $\phi_C$ and $\phi_C^{\ell\ell}$ distributions for both the CP-even and the CP-odd Higgs bosons, we are ready to discuss how well one can distinguish the CP-odd Higgs boson from the CP-even one. In our study we divide the $\phi_C$ and $\phi_C^{\ell\ell}$ distributions into 10 bins and use the binned likelihood function defined as following:
\beq
L(\mu,\alpha)\equiv\prod_{i=1}^{N_{\text{bin}}}\frac{\left(\mu s_i(\alpha)+b_i\right)^{n_i}}{n_i!}e^{-\mu s_i(\alpha)-b_i},
\eeq
where $N_{\text{bin}}=10$, $\mu$ is the strength of the signal, $b_i$ and $n_i$ is the number of the background and observed event in the $i$th bin, respectively, and $s_i(\alpha)$ is the number of the signal event in the $i$th bin for the CP phase $\alpha$. 

\begin{figure}
\begin{center}
  \includegraphics[width=0.45\textwidth]{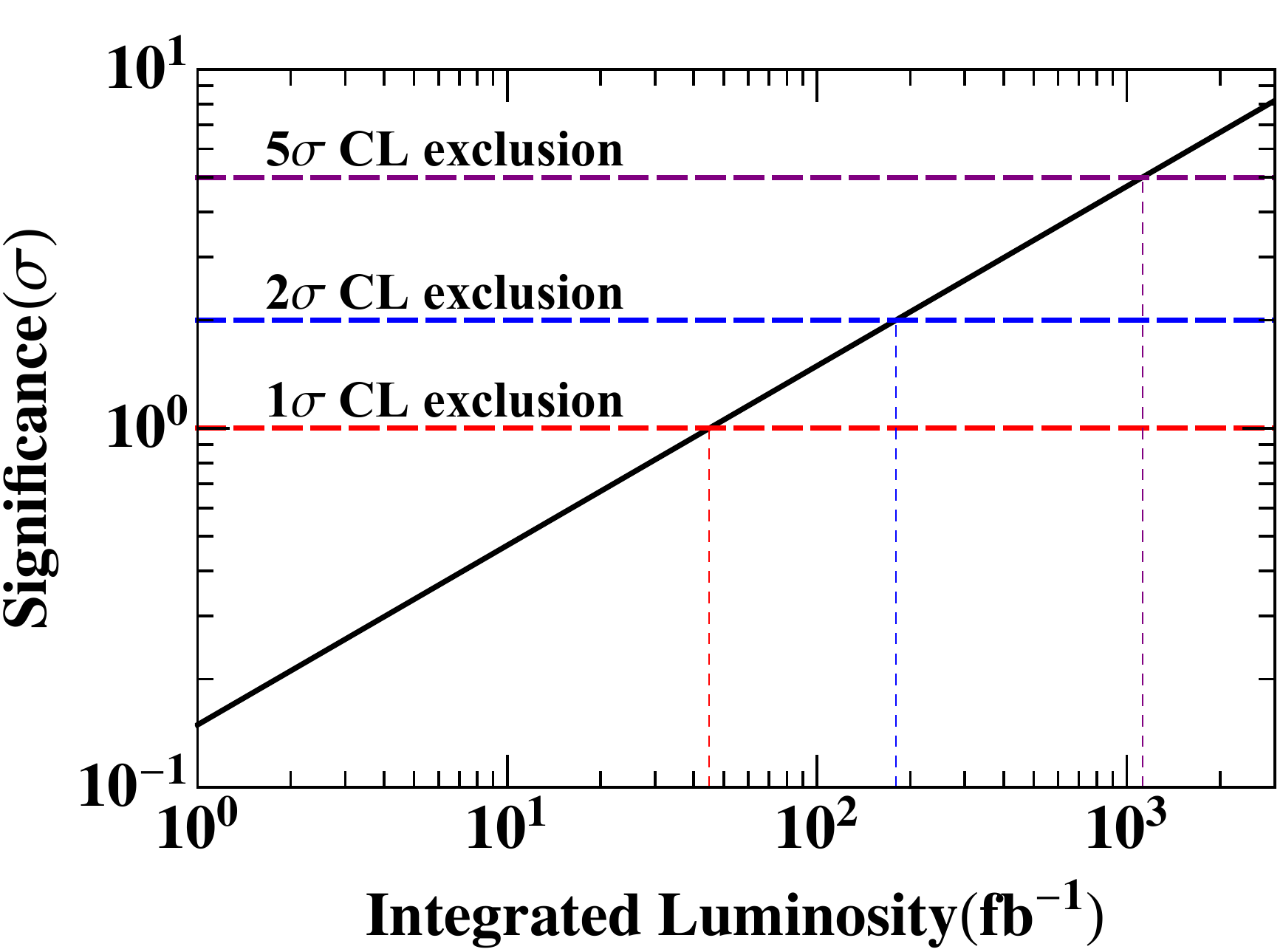}
\caption{The statistical significance of discriminating the CP-odd Higgs boson from the CP-even Higgs boson as a function of the integrated luminosity at 14 TeV LHC. 
}
\label{fig6}
\end{center}
\end{figure}

The recent measurement of the $t\bar{t}H$ production shows that the signal event number is in consistent with the SM prediction~\cite{Sirunyan:2018koj,Aaboud:2018urx}. We thus rescale $\mu$ for all the $\alpha$'s to match the signal strength of the SM value. The logarithm of likelihood function ratio is defined as 
\beq
-2\log\lambda(\alpha_1|\alpha_0)=-2\log\frac{L(\hat\mu_1,\alpha_1)}{L(\hat\mu_0,\alpha_0)},
\label{eq:llr}
\eeq
where $\hat\mu_k~(k=0,1)$ is determined by minimizing $-2\log L(\hat\mu_k,\alpha_k)$. Setting $n_i=\hat\mu_0s(\alpha_0)_i+b_i$, the hypothesis 1 is excluded versus the hypothesis 0 with $\sqrt{-2\log\lambda(\alpha_1|\alpha_0)}\sigma$ confidence level (CL). Using this relation, we combine the diphoton and the $b\bar b$ channels to obtain the statistic significance of distinguishing a CP-odd Higgs boson from a CP-even Higgs boson.  Figure~\ref{fig6} displays the exclusion significance as a function of the integrated luminosity at 14 TeV LHC.  It shows that, if the Higgs boson is a pure CP-even scalar, in order to exclude the pure CP-odd hypothesis at 95\% CL, one expects that an integrated luminosity of $\sim 180~\rm{fb}^{-1}$ will be needed. 

\noindent {\bf 4. Measurement of the CP-phase angle $\alpha$}
\vspace{2mm}

\begin{figure}
\begin{center}
 \includegraphics[width=0.4\textwidth]{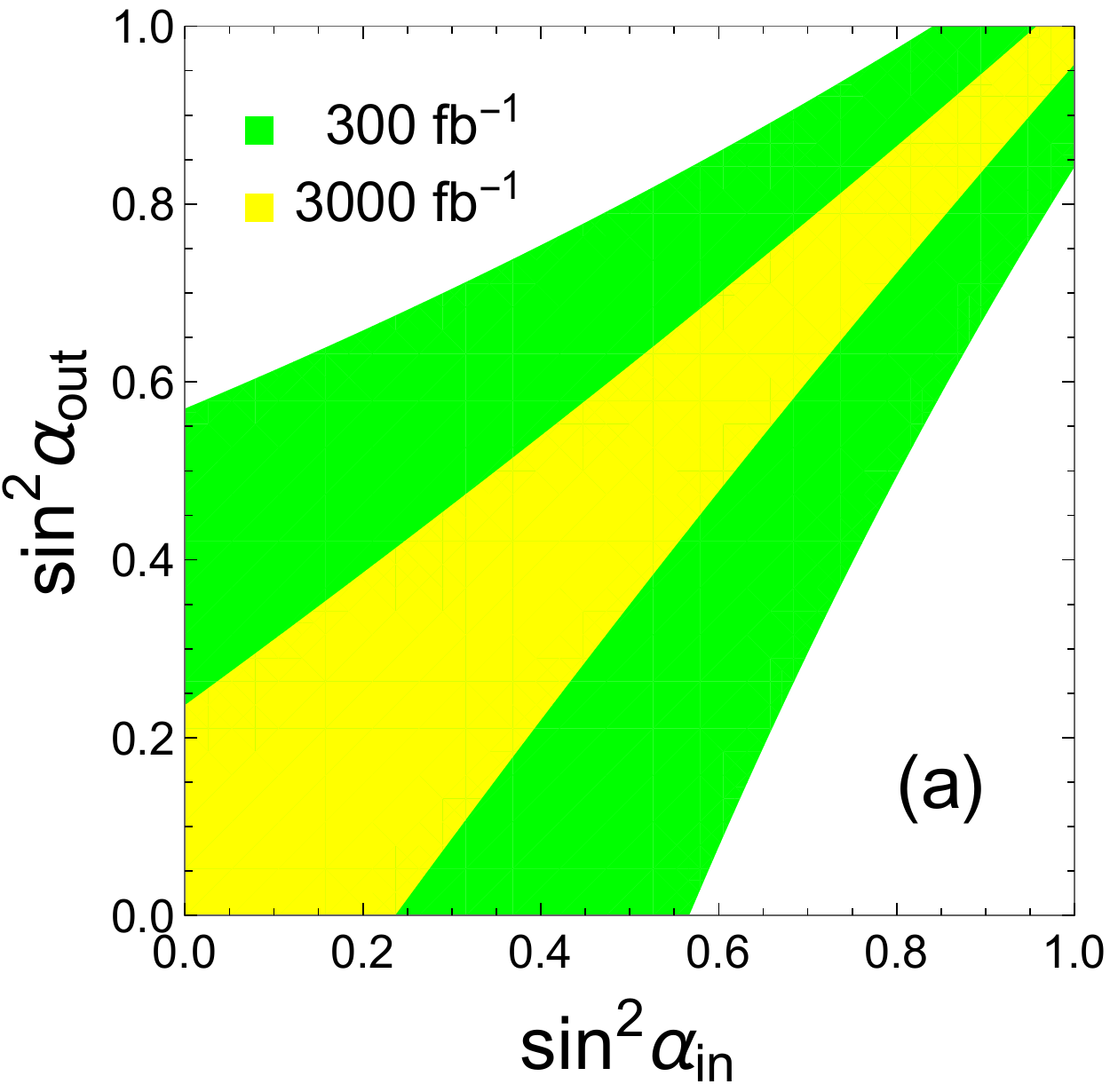} \\
 \includegraphics[width=0.4\textwidth]{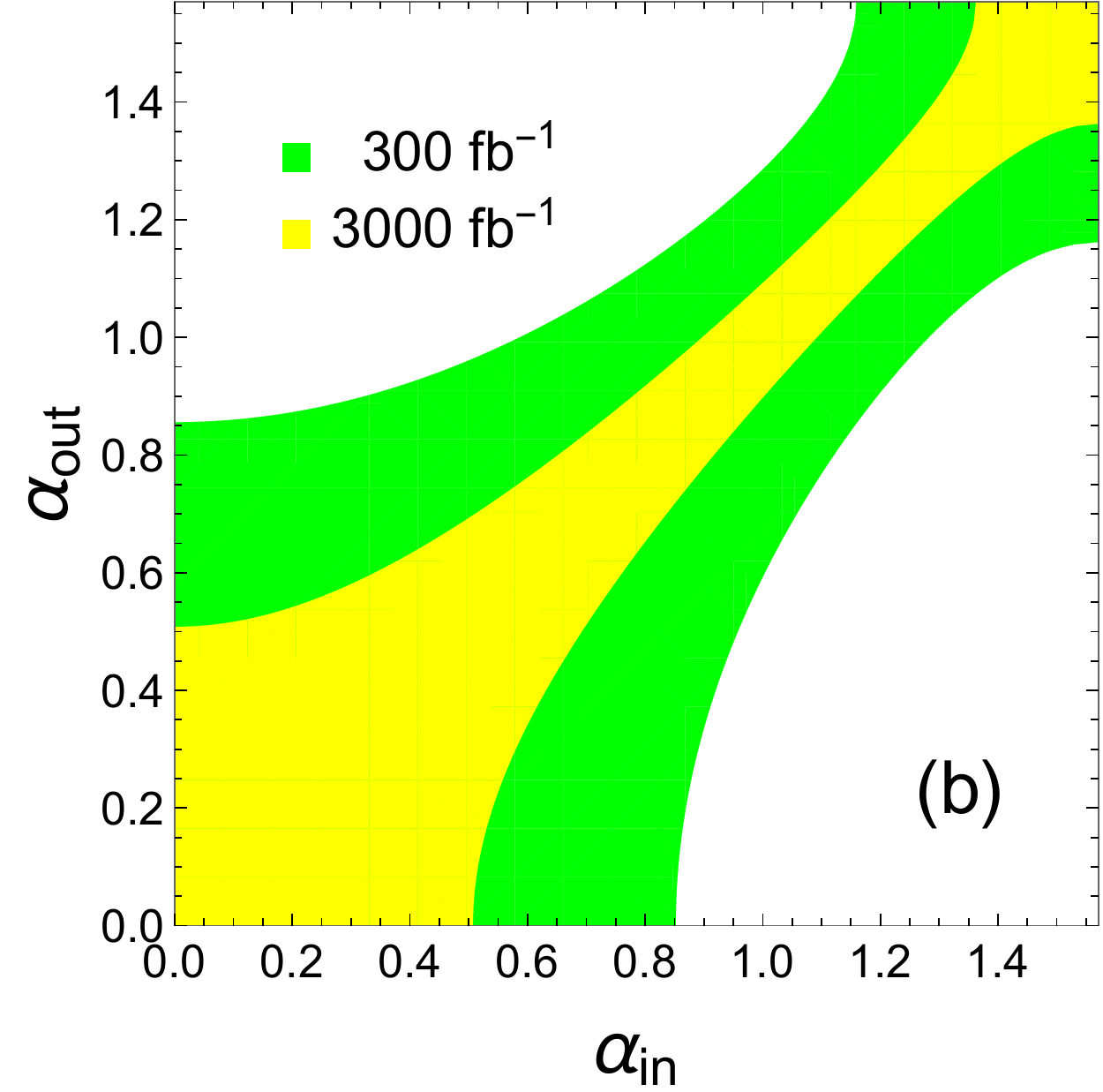}
\caption{The projected accuracy of the $\alpha$ measurement versus the input value at the LHC with an integrated luminosity of $300~{\rm fb}^{-1}$ (green) and $3000~{\rm fb}^{-1}$ (yellow), respectively. 
}
\label{fig7}
\end{center}
\end{figure}

Now we discuss how well one can measure the CP-phase angle $\alpha$ from the $\phi_C$ and $\phi_C^{\ell\ell}$ distributions. In general, the $\phi_C$ and $\phi_C^{\ell\ell}$  distributions of the signal channel can be written as 
\beq
s(\alpha)=A\cos^2\alpha+B\cos\alpha\sin\alpha+C\sin^2\alpha.
\label{eq:signal}
\eeq
Note that the $A$ and $C$ term corresponds to the CP-even and CP-odd contribution, respectively, and the $B$ term is zero for the $t\bar{t}H$ production. After dividing the $\phi_C$ and $\phi_C^{\ell\ell}$ distributions into 10 bins, we read out the CP-even ($\alpha=0$) and the CP-odd ($\alpha=\pi/2$) contribution in each bin, defined as $s_i(0)$ and $s_i(\pi/2)$, respectively. Therefore, the distribution of the signal event is given by 
\beq
s_i(\alpha)=s_i(0)\cos^2\alpha+s_i(\pi/2)\sin^2\alpha.
\eeq
The $t\bar{t}H$ production has been confirmed recently by both the ATLAS and CMS collaborations, assuming a purely CP-even Higgs boson~\cite{Sirunyan:2018koj,Aaboud:2018urx}. The current data of the signal strength, $\mu=1.18^{+0.30}_{-0.27}$~\cite{Sirunyan:2018koj}, is well consistent with the SM theory though it has a large experimental uncertainty. To explore the potential of measuring the angle $\alpha$ in future experiments, we rescale the signal strength $\mu$ of the input angle $\alpha$ to be the same as the SM theoretical prediction. 

We vary the signal strength $\mu$ for each input $\alpha$ to minimize the logarithm of likelihood function ratio (the signal strength which minimizes the $-2\log L(\mu,\alpha)$ is denoted as $\hat\mu$ here), defined as 
\beq
-2\log\lambda(\alpha;\alpha_0)=-2\log\frac{L(\hat\mu,\alpha)}{L(\hat\mu_0,\alpha_0)},
\eeq
to obtain the projected sensitivity of the $\alpha$ measurement. The following condition, 
\begin{equation*}
-2\log\lambda(\alpha;\alpha_0)\leqslant1,
\end{equation*}
yields the $1\sigma$ confidence interval of the measured $\alpha$ angle for a given input $\alpha_0$. As shown in Eq.~(\ref{eq:signal}), the signal rate depends on $\sin^2\alpha$ rather than directly on $\alpha$; therefore, we first obtain the sensitivity of the future LHC experiment on $\sin^2\alpha$. Figure~\ref{fig7}(a) displays the projected experimental measurement of $\sin^2\alpha_{\rm out}$ versus the theoretical input $\sin^2\alpha_{\rm in}$ at 14~TeV LHC with an integrated luminosity of $300~{\rm fb}^{-1}$ (green) and $3000~{\rm fb}^{-1}$ (yellow), respectively.  The uncertainty of the $\sin^2\alpha$ measurement is large in the region of $\alpha\sim 0$ and is reduced in the region $\alpha \sim \pi/2$. Increasing the integrated luminosity significantly reduces the uncertainties; see the yellow band. Figure~\ref{fig7}(b) shows the correlation between the $\alpha_{\rm out}$ and $\alpha_{\rm in}$. Owing to the small production rate, it is still very challenging to achieve a precise knowledge of the CP-phase $\alpha$ at the high-luminosity LHC. 

The behavior of the contours can be qualitatively understood as follows. From the definition of the likelihood ratio given in Eq.~(\ref{eq:llr}), it is easy to show that 
\bea
-2\log\lambda(\alpha;\alpha_0)&=&2\sum_{i=1}^{N_{\rm bin}}\biggl\{\hat\mu s_i(\alpha)-\hat\mu_0s_i(\alpha_0)\nonumber\\
&&\left.-n_i \log\left[1+\frac{\hat\mu s_i(\alpha)-\hat\mu_0 s_i(\alpha_0)}{n_i}\right]\right\},~~~~~
\eea
where $n_i=\hat\mu_0s_i(\alpha_0)+b_i$. We demand the number of the signal event to be the same as the SM case in order to respect the current data. As a result, it yields
\beq
\sum_{i=1}^{N_{\rm bin}}\biggl(\hat\mu s_i(\alpha)-\hat\mu_0s_i(\alpha_0)\biggr)=0.
\eeq
Note that the above condition is valid only after summing over all the bins. Using a rough approximation of each bin,
\beq
\big|\hat\mu s_i(\alpha)-\hat\mu_0 s_i(\alpha_0)\big|<n_i,
\eeq
we expand the logarithm of likelihood ratio function to the second order and obtain 
\bea
-2\log\lambda(\alpha;\alpha_0)\approx\sum_i\frac{[\hat\mu s_i(\alpha)-\hat\mu_0 s_i(\alpha_0)]^2}{\hat\mu_0s_i(\alpha_0)+b_i}.
\label{eq:18}
\eea
By definition $\hat\mu_0=1$ when $\alpha_0=0$, and it yields
\beq
\hat\mu_0=\left[\cos^2\alpha_0+\frac{\sum_is_i(\pi/2)}{\sum_is_i(0)}\sin^2\alpha_0\right]^{-1}.
\eeq
Through simple algebra one can show that 
\[ -2\log\lambda(\alpha;\alpha_0)\propto \sin^22\alpha_0,\]
which explains the linear behavior of the contour in Fig.~\ref{fig7}(a).

\vspace{6mm}
\noindent{\bf 5. Conclusions and Discussion}
\vspace{2mm}

We proposed a new observable $\phi_C$ to measure the CP property of the top quark Yukawa coupling in the $t\bar t H$ production. The observable $\phi_C$ is the  dihedral angle between the plane of the incoming protons and the plane of the top quark pair in the rest frame of the Higgs boson. We carry out a fast simulation of the $t\bar{t}H$ production with two decay modes of the Higgs boson, $H\to b\bar{b}$ and $H\to \gamma\gamma$, and the SM background process of $t\bar{t}\gamma\gamma$. Both the CP-even $Ht\bar{t}$ coupling and the SM background process of $t\bar{t}\gamma\gamma$ has similar shape in the $\phi_C$ distribution before and after the kinematic cuts. On the other hand, the CP-odd coupling exhibits different $\phi_C$ distribution such that it serves well for searching for the CP-odd coupling. At the 14~TeV LHC with an integrated luminosity of $\sim 180~\mbox{fb}^{-1}$ one can distinguish the CP-odd coupling from the CP-even hypothesis at the 95\% confidence level.  

\vspace{6mm}
\noindent{\bf Acknowledgement:}
QHC and RZ are supported in part by the National Science Foundation of China under Grant Nos. 11725520, 11675002, 11635001. KPX is supported by grant NRF-2019R1C1C1010050. HZ is supported by Institute of High Energy Physics, Chinese Academy of Science, under Contract No. Y6515580U1 and Innovation Grant Contract No. Y4545171Y2.

\bibliographystyle{apsrev}
\bibliography{reference}
  
\end{document}